%% file: main.tex
\documentclass[]{interact}

\usepackage{epstopdf}% To incorporate .eps illustrations using PDFLaTeX, etc.
\usepackage{subfigure}% Support for small, `subfigures and tables
\usepackage{array}
\usepackage{graphicx}
\usepackage{float}

\usepackage{natbib}% Citation support using natbib.sty
\bibpunct[, ]{(}{)}{;}{a}{}{,}% Citation support using natbib.sty
\renewcommand\bibfont{\fontsize{10}{12}\selectfont}% Bibliography support using natbib.sty

\theoremstyle{plain}% Theorem-like structures provided by asthma. sty

\theoremstyle{definition}

\theoremstyle{remark}

\begin{document}

\articletype{ARTICLE TEMPLATE}

\title{Unleashing the Power of AI: Transforming Marketing Decision-Making in Heavy Machinery with Machine Learning, Radar Chart Simulation, and Markov Chain Analysis}
\author{
    Tian Tian\\
    Stuart Business School, Illinois Institute of Technology \\
    \texttt{ttian4@hawk.iit.edu},    \vspace{10pt} \\
    Jiahao Deng\\
    College of Computing and Digital Media, DePaul University \\
    \texttt{jdeng5@depaul.edu}
    \vspace{10pt} 
}

\maketitle
\author{
    Tian Tian\textsuperscript{a}\thanks{CONTACT A.~N. Author. Email: latex.helpdesk@tandf.co.uk} 
    Jiahao Deng\textsuperscript{b}  Kai Shu\textsuperscript{a}   \\
     \vspace{10pt} \\
    \textsuperscript{a}Stuart Business School, Illinois Institute of Technology,
    \texttt{ttian4@hawk.iit.edu}\\ 
    \textsuperscript{b}College of Computing and Digital Media, DePaul University, \texttt{jdeng5@depaul.edu} }

\begin{abstract}
This pioneering research introduces a novel approach for decision-makers in the heavy machinery industry, specifically focusing on production management. The study integrates machine learning techniques like Ridge Regression, Markov chain analysis, and radar charts to optimize North American Crawler Cranes market production processes. Ridge Regression enables growth pattern identification and performance assessment, facilitating comparisons and addressing industry challenges. Markov chain analysis evaluates risk factors, aiding in informed decision-making and risk management. Radar charts simulate benchmark product designs, enabling data-driven decisions for production optimization. This interdisciplinary approach equips decision-makers with transformative insights, enhancing competitiveness in the heavy machinery industry and beyond. By leveraging these techniques, companies can revolutionize their production management strategies, driving success in diverse markets.
\end{abstract}


\input{Introduction}
\input{Approach}

\input{Benchmark}
\input{Stochastic}
\input{Conclusion}

%\section*{Acknowledgement(s)}
%An unnumbered section, e.g.\ \verb"\section*{Acknowledgements}", may be used for thanks, etc.\ if required and included \emph{in the non-anonymous version} before any Notes or References.

\section*{Disclosure statement}
The authors report there are no competing interests to declare.

%\section*{Funding}

%An unnumbered section, e.g.\ \verb"\section*{Funding}", may be used for grant details, etc.\ if required and included \emph{in the non-anonymous version} before any Notes or References.

%\section*{Notes on contributor(s)}

%An unnumbered section, e.g.\ \verb"\section*{Notes on contributors}", may be included \emph{in the non-anonymous version} if required. A photograph may be added if requested.

%\section*{Nomenclature/Notation}

%An unnumbered section, e.g.\ \verb"\section*{Nomenclature}" (or \verb"\section*{Notation}"), may be included if required, before any Notes or References.

%\section*{Notes}

%An unnumbered `Notes' section may be included before the References (if using the \verb"endnotes" package, use the command \verb"\theendnotes" where the notes are to appear, instead of creating a \verb"\section*").

%\begin{verbatim}
\bibliographystyle{tfcad}
\bibliography{refs.bib}
%\end{verbatim}

\end{document}

%% file: Introduction.tex
\section{Introduction}
This pioneering research significantly contributes to the industrial management of marketing decision-making in the heavy machinery industry by combining machine learning, data analysis, traditional Markov chain, and radar chart visualization. In addition, the findings and methodologies of this study have broader applications and can benefit decision-makers. Integrating data-driven methods, such as machine learning, data analysis, Markov chain statistics, and radar charts, offers valuable insights, risk assessment capabilities, and visualization tools that decision-makers in various industries can harness to optimize performance, manage risks, and drive strategic decision-making.
\subsection{Introduction of data-driven decision-making and risk factors}
In today's dynamic and competitive business landscape, data-driven decision-making has become essential for marketing professionals to gain a competitive edge. The heavy machinery industry, in particular, requires meticulous analysis of market dynamics, competition, and risk factors to make informed strategic choices. A data-driven decision-making study aims to identify the competitive landscape, understand the market dynamics, and assess the strategies of key players in the industry (\cite{8,9}). Data-driven decision-making study in the heavy machinery industry involves evaluating the strengths and weaknesses of key business players, including comparing manufacturers' revenue, revenue size, revenue growth rate, and competition relationship, which are essential factors to consider (\cite{10,11,12}). By conducting a competitive business strategy analysis, companies in the heavy machinery industry can identify their strengths and weaknesses, benchmark themselves against their competitors, and make informed strategic decisions to improve their competitiveness and business position (\cite{13,14,15}).
The revenue comparison helps understand how each manufacturer performs in the market and identifies the leading players (\cite{16,17}). It is also crucial to evaluate the revenue size of each manufacturer to determine their market share and dominance.
Furthermore, analyzing the revenue growth rate of each manufacturer can provide insight into their future growth potential and their ability to adapt to changing market conditions (\cite{18,19,20}). A higher revenue growth rate indicates a stronger market position and potential for future success (\cite{21}).
Finally, understanding the competition relationship between manufacturers is crucial for identifying potential threats and opportunities (\cite{22}). This can be achieved by analyzing pricing, product offerings, distribution channels, and marketing strategies.
Overall, conducting a thorough analysis of these factors can help manufacturers make informed decisions about their business strategies and improve their competitiveness in the heavy machinery equipment market.

\subsection{Trend analysis}
This paper explores the innovative application and benefits of utilizing Ridge Regression as a powerful tool for trend analysis in business and industrial management, focusing on the competitive market for crawler cranes in the heavy machinery sector. Understanding and predicting market trends is paramount for effective decision-making and maintaining a competitive edge in the industrial machinery industry, especially within the crawler crane segment. Traditional regression methods often grapple with challenges related to multicollinearity and overfitting when analyzing intricate datasets (\cite{53}). Nevertheless, Ridge Regression rises to the occasion by introducing a regularization parameter that mitigates model complexity and stabilizes coefficient estimates.

The novelty of integrating Ridge Regression into trend analysis lies in its capacity to strike a harmonious balance between bias and variance (\cite{55}). By imposing a penalty term on the regression model, Ridge Regression adeptly manages the influence of highly correlated predictors, resulting in more precise and resilient trend analysis (\cite{57}). This approach equips industrial management decision-makers with the tools to identify and interpret pivotal factors influencing market trends, facilitating proactive strategies and well-informed actions.

Furthermore, the advantages of employing Ridge Regression for trend analysis in the heavy machinery crawler crane market are substantial. Firstly, it empowers decision-makers to discern critical variables that shape market trends, such as customer preferences, technological advancements, and regulatory changes (\cite{59}). With this knowledge, companies can align their product development, marketing strategies, and resource allocation with these identified trends, ultimately enhancing competitiveness and market positioning.

Secondly, Ridge Regression quantifies the impact of each predictor on market trends (\cite{61}). Consequently, decision-makers can prioritize their efforts based on the magnitude of influence and allocate resources judiciously. This approach streamlines decision-making and resource allocation, optimizing overall business and industrial management performance in the fiercely competitive heavy machinery market.

\subsection{Risk factors analysis}
Integrating advanced analytical models into marketing analysis has ushered in a transformation in how businesses formulate strategic decisions (\cite{1, 2, 3}). While many techniques have found application in this domain, using Markov chains, a mathematical model historically rooted in physics and finance, remains relatively uncommon in business and industrial management (\cite{4}). However, the distinct capabilities inherent in Markov chains hold significant potential to unearth invaluable insights and revolutionize the decision-making processes in the heavy machinery industry.

Markov chains present a probabilistic framework that equips marketing professionals with the means to model and comprehend the intricate dynamics of complex systems (\cite{5, 6}). At its core, the Markov chain hinges on its capacity to predict future states of a system based solely on its present form, with no consideration of past events. This unique property renders it an ideal instrument for scrutinizing competitive relationships, assessing company performance, and evaluating risk factors within the heavy machinery industry.

By harnessing the power of the Markov chain model, decision-makers in marketing can discern patterns, trends, and probabilities associated with future events in the marketplace (\cite{7}). This predictive prowess empowers them to make well-informed decisions and craft effective marketing strategies that propel business growth and enhance their competitive standing.

This research highlights Markov chains' underexplored yet promising application in marketing analysis within business and industrial management, particularly in the heavy machinery industry. By adopting the Markov chain model, decision-makers can comprehensively understand market dynamics, pinpoint potential risks, and base their decisions on data-driven insights to optimize their marketing endeavors.

This study serves as a clarion call to recognize the untapped potential of Markov chains in marketing analysis. It underscores their pertinence in the sphere of business and industrial management. Furthermore, by showcasing the tangible benefits of this approach and providing practical insights, we seek to inspire further exploration and the broader adoption of the Markov chain model in the decision-making processes of marketing professionals. Ultimately, we aim to empower marketing practitioners in the heavy machinery industry to harness advanced analytical techniques, unlocking new vistas of opportunity and success.

\subsection{Radar chart to simulate benchmark products}
Using a radar chart to manufacture benchmark products offers several advantages and novel applications (\cite{54,64}):

\textbf{Visual Comparison}: Radar charts provide a visually appealing and intuitive way to compare multiple attributes or dimensions of different products simultaneously. The radar chart allows for a quick and comprehensive visual comparison of the benchmark products by plotting the details on different axes and connecting the data points (\cite{56,65}).

\textbf{Comprehensive Assessment}: Radar charts thoroughly assess the benchmark products across various dimensions. Each attribute represents a specific characteristic or feature of the product, such as quality, performance, features, price, or customer support. The radar simultaneously hart provides a holistic view of how the benchmark products permission by plotting the data points for each attribute (\cite{58,66}).

\textbf{Relative Performance}: Radar charts enable the assessment of the benchmark products' relative performance compared to each other. The distance from the center of the chart to each data point represents the magnitude or value of the attribute. By comparing the spaces between the data points of different products, it is possible to identify which product performs better or worse in specific dimensions (\cite{60,67}).

\textbf{Identifying Strengths and Weaknesses}: Radar charts help identify the strengths and weaknesses of the benchmark products across different dimensions. Observing the patterns and shapes formed by the data points makes it easier to identify areas where a product excels or lags. This information can be invaluable for decision-making and prioritizing improvements or investments in specific areas (\cite{62,68}).

\textbf{Visualization of Trade-offs}: Radar charts allow for the visualization of trade-offs between different attributes or dimensions. When the data points of two or more products intersect or overlap, they indicate that they have similar performance levels in those dimensions. On the other hand, when the data points diverge, it highlights the trade-offs or differences between the products (\cite{63,69}).

Overall, using a radar chart to simulate benchmark products offers a novel and effective way to visually compare and assess the performance of different products across multiple dimensions (\cite{70}). It facilitates a comprehensive understanding of the product's strengths, weaknesses, and trade-offs, enabling informed decision-making and strategic planning. The approach stated in section 2; section 3 gives Benchmark product simulation and risk factors extraction. Section 4 is a factors analysis incorporating the Markov process. Finally,  section 5 is the future study of Brownian motion and the paper's conclusion. 

%% file: Approach.tex
\section{Approach}
\subsection{Data}
This research used the dataset from third-party vendors, including 5-year (2017-2021)  revenues from 8 companies in the North America Crawler Cranes market. Tadano Ltd, Zoomllon Heavy Industry Science and Technology Co. Ltd, Sany Heavy Industry Co. Ltd, Kato Works Co. LTD, Liebherr, Terex Corporation, Altec Industries, etc.  The dataset is relatively small for the following analysis, so random noise was added to help increase the data set size and reduce the risk of over-fitting. 
\subsection{Process of analysis}
To compare the products of Crawler Cranes and their competitiveness relationship among companies Tadano Ltd, Zoomllon Heavy Industry Science and Technology Co. Ltd, Sany Heavy Industry Co. Ltd, Kato Works Co. LTD, Liebherr, Terex Corporation, Altec Industries, and other Crawler Cranes manufacturers, this article provides a set of analysis of revenues, growth rate, and competition relationship 
 starting with illustrating the background of Crawler Crane Market with the introduction of essential market performance, followed by collecting data of payments from 7 corporate in the Crawler Cranes market from 2017 to 2021 from a third-party data vendor. Then, preprocessed the data by adding random noise since the dataset is relatively small for analysis. Visualization, including histograms, boxplots, line charts, etc., helped me understand the data distribution and hidden patterns. Then, correlation heatmap and clustering show the competition relationships between companies. 
\\
\textbf{Step 1}: Standard Normalization and Random Noise Addition
Let $X$ be the original data matrix of size $N \times T$, where  $x_{i,t}$ is the revenue of company I at time t. To normalize the data, we apply standard normalization:
$Z_{i,t}=\left(x_{i,t}-\mu_i\right)/\sigma_i$
where $\mu_i$ and $\sigma_i$  are the mean and standard deviation of company I over all periods.
To add random noise, we add a matrix E of size N X T, where each element $e_{i,t}$
 is a random value drawn from a normal distribution with mean 0 and standard deviation $\delta$? The normalized and noisy data matrix is then given by:
$Y_{i,t}=Z_{i,t}+e_{i,t}$
\\
\textbf{Step 2}: Clustering Analysis: 
Let D be the pairwise distance matrix between companies, where $D_{i,j}$ represents the distance between companies i and j. Then, the clustering algorithm assigns each company I to a cluster $c_i$, minimizing the sum of the pairwise distances between companies in the same group. This objective function can be written as:

minimize $\sum_{k=1}^{K}\sum_{i,j\in C_k}D_{i,j}$
subject to $\bigcup_{k=1}^KC_k=1,2,\ldots,N$

Where K is the number of clusters, N is the total number of companies, $C_k$ is the set of companies assigned to cluster k, and the second constraint ensures that each company belongs to exactly one cluster.
\\
\textbf{Step 3}: Correlation Analysis
To generate a correlation map, we calculate the Pearson correlation coefficient between the revenue time series of each pair of companies:

$r_{i,j}=corr\left(Y_i,Y_j\right)=cov\left(Y_i,Y_j\right)/\left(\sigma_i\ast\sigma_j\right)$
\\
Where $cov\left(Y_, Y_j\right)$ is the covariance between the revenue time series of companies i and j, and $\sigma_i$ and $\, sigma_j$ are their respective standard deviations. The correlation coefficients can be visualized as a heatmap to show the strength and direction of correlations between the companies.

\subsection{Results}
\subsubsection{Revenue Size and Growth Rate}
\begin{figure}[t]
\centering
\includegraphics[width=12cm]{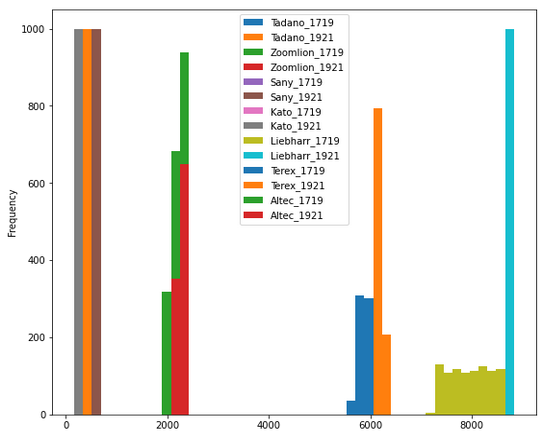}
\caption{\centering North America Crawler Cranes Market Revenue
}
\label{fig:2.1}
\end{figure}

The frequency distribution chart (Figure \ref{fig:2.1}) provides information about revenue distribution among North American crawler crane market companies. There are three clusters, with the most excellent payment having a significant advantage over the others. It also shows that Liebherr has a significantly larger market share than any other company during the five years, with Terex following closely behind. Additionally, the chart indicates that Zoomlion has a slightly higher market share than Altec, while Tadano and Sany are relatively close in revenue from 2017 to 2019. Moreover, from 2019 to 2021, the top two companies will remain dominant, and the third and fourth revenue firms will likely face increased competition. Additionally, it is worth mentioning that Kato has the lowest revenue share compared to the other six firms from 2019 to 2021. Moreover, the chart indicates that the highest revenue cluster includes Liebherr and Terex, while the group with medium revenue includes Zoomlion and Altec. The low-revenue collection includes Tanado, Sany, and Kato.

\begin{figure}[t]
\centering
\includegraphics[width=12cm]{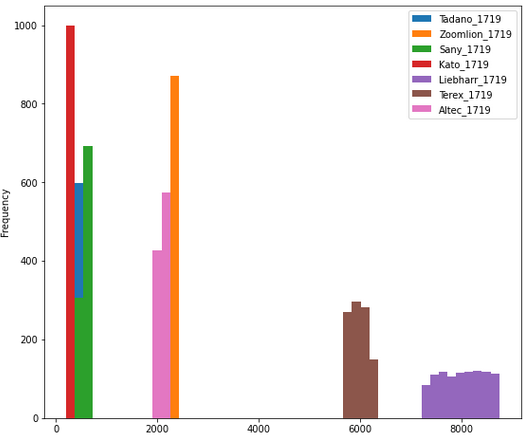}
\caption{\centering Revenue Growth Rate Distribution from 2017 to 2019}
\label{fig:2.2}
\end{figure}

The chart (Figure \ref{fig:2.2}) shows the growth rate distribution from 2017 to 2019. According to the frequency chart, Liebherr has a tremendous growth rate from 2017 to 2019. Terex has also seen a quite substantial growth rate which is around $10\%$ between 2017 and 2019. Except for these two companies, other companies have slow growth rate and some even have a declining revenue during this time period. 

\begin{figure}[t]
\centering
\includegraphics[width=12cm]{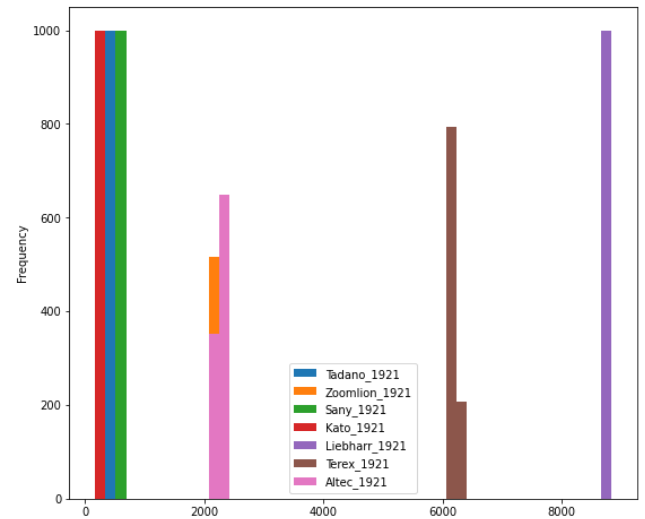}
\caption{\centering Revenue Growth Distribution from 2019 to 2021
}
\label{fig:2.3}
\end{figure}

The Figure \ref{fig:2.3} shows the revenue growth distribution from 2019 to 2021. In this period, the gap between Liebherr's revenue and other companies' revenue becomes more pronounced between 2019 and 2021, and the revenue difference between Liebherr and Terex is more evident in this time period as well. One thing needs to be mention is that the revenue from crawler cranes barely varied across all companies during 2019 to 2021, and attribute this to the impact of the 2020 pandemic. This suggests that the pandemic may have had a significant effect on the revenue generated by companies in this market.

\begin{figure}[t]
\centering
\includegraphics[width=12cm]{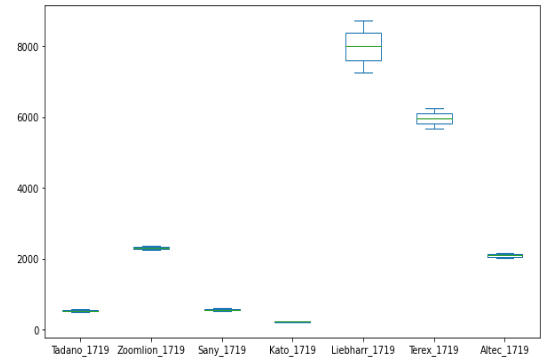}
\caption{\centering Boxplot of Revenue Distribution from 2017 to 2019
}
\label{fig:2.4}
\end{figure}

Compared to the other six firms, the distribution study (Figure \ref{fig:2.4}) reveals that Liebherr had the highest revenue and the quickest growth rate from 2017 to 2019. Terex ranked second in revenue in crawler crane market and grew significantly as well from 2017 to 2019. Moreover, Altec was expanding more quickly than Zoomlion and other companies. 

\begin{figure}[t]
\centering
\includegraphics[width=12cm]{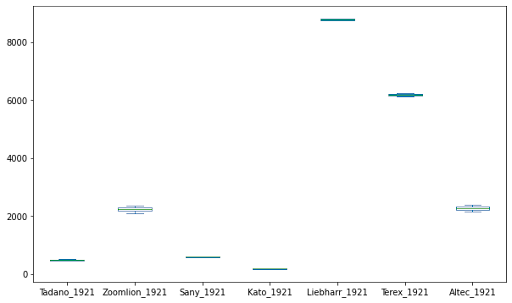}
\caption{\centering Boxplot of Revenue Distribution from 2019 to 2020
}
\label{fig:2.5}
\end{figure}

\begin{figure}[t]
\centering
\includegraphics[width=12cm]{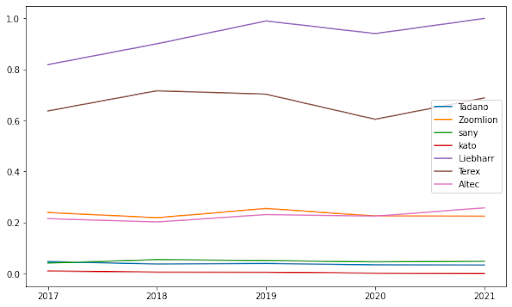}
\caption{\centering Growth Analysis of Revenue from 2017 to 2021
}
\label{fig:2.6}
\end{figure}

On the contrary, from 2019 to 2021 all companies were seen only modest revenue growth rate according to Figure \ref{fig:2.5} Also, some companies experienced a decline in revenue in 2020 mainly due to the pandemic. 
The Line chart (Figure \ref{fig:2.6}) confirms the results of the previous graphical analysis. The data is normalized using for clear presentation. In conclusion, Liebherr’s revenue increased significantly between 2017 to 2019. Terex has the second-highest rate of revenue growth from 2017 to 2019. However, between 2019 and 2021, all corporate are expected to develop slowly, which the pandemic might impact. 

\subsubsection{Competition Relationship}
The heatmap shows the competition relationship between crawler cranes companies. Since 2017 to 2019 and 2019 to 2021 have significant revenue differences due to the pandemic, the analysis divided the dataset into two clusters by period from 2017 to 2019.

\begin{figure}[t]
\centering
\includegraphics[width=12cm]{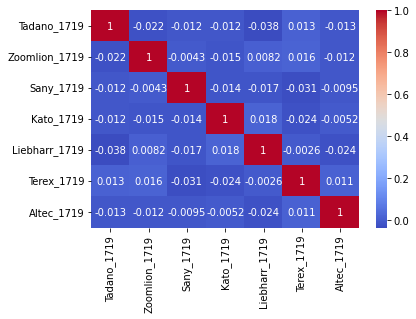}
\caption{\centering Heatmap of Competition Relationship from 2017 to 2019
}
\label{fig:2.7}
\end{figure}

The heatmap (Figure \ref{fig:2.7}) provides information about the correlation between the revenue generated by different companies in the North American crawler crane market. Tadano and Altec are comparable in this analysis from 2017 to 2019 and have a positive connection with Terex and an opposite direction of revenue movement with the other five firms. The revenue trend for Liebherr is the same as for Zoomlion and Kato. In the correlation table, the positive correlation with Kato has the most significant positive value (0.18). In addition, despite Liebherr having a relatively modest positive correlation parameter with Zoomlion, Liebherr's fast growth rate from 2017 to 2019 also benefitted from the positive correlation with Zoomlion, a medium-sized business. Also, Zoomlion's revenue is positively correlated with the gain of only two companies, both of which are large companies, including Liebherr and Terex. In contrast, Sany has a negative correlation with all other companies, and Kato has a negative correlation with almost all companies except Liebherr. Regarding Terex, it had a negative correlation with half of the firms and a positive correlation with the other half, and the positive correlation's parameter is relatively strong. This is probably because Terex's revenue increased significantly from 2017 to 2019. 

\begin{figure}[t]
\centering
\includegraphics[width=12cm]{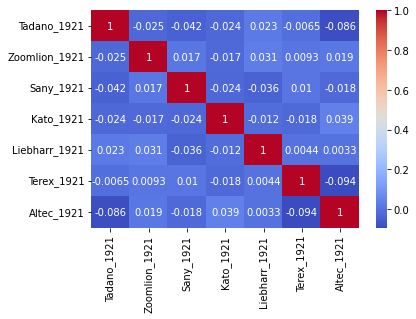}
\caption{\centering Heatmap of Competition Relationship from 2019 to 2021
}
\label{fig:2.8}
\end{figure}

\begin{figure}[t]
\centering
\includegraphics[width=12cm]{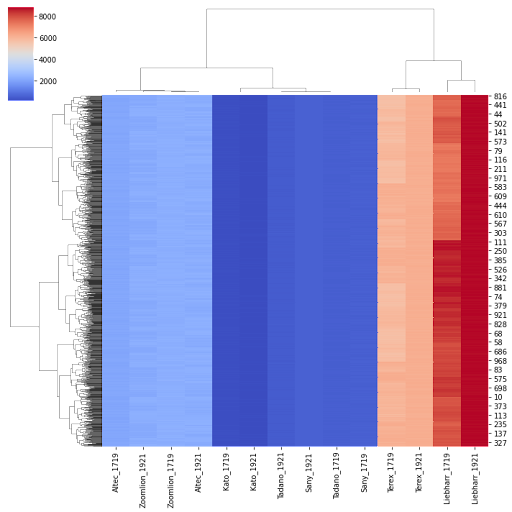}
\caption{\centering Cluster Analysis of Revenue in the period 2017 to 2019 and 2019 to 2021
}
\label{fig:2.9}
\end{figure}

From 2019 to 2021, the situation is slightly different. Terex maintained a negative correlation with three companies and a positive correlation with three. Still, the negative numbers were large in absolute terms this time, especially with Altec's negative parameters. As a result, it may intensify the drop in Terex's revenues from 2019 to 2021.
From the previous positive association with Terex to the current positive association with Liebherr alone, Tadano has changed. However, Zoomlion, Sany, Kato, and especially Altec have a more profound negative association with Tadano than from 2017 to 2019. Figure \ref{fig:2.8} shows that Zoomlion has a favorable revenue associated with most firms from 2019 to 2021 while only negatively correlating with Tadano and Kato. Kato continues to have a positive correlation with only one company these years. This seems to explain the decrease in its revenue from 2019 to 2021. Altec has the most significant positive correlation magnitude with Kato, and the two most enormous absolute negative numbers correlate with Tadano and Terex.  
According to the cluster analysis (Figure \ref{fig:2.9}), Liebherr and Terex's revenues have a stronger correlation with their historical revenues. Zoomlion and Altec have a stronger correlation in their payments over the same period. The remaining three firms' sales are more closely correlated. Also, it confirms that the two time periods had significantly different revenues among the entire industry. 

\subsection{Trend Analysis}
Performing a trend analysis on the yearly sales data for the seven companies is essential because it allows us to identify and quantify the underlying growth patterns and compare the performance of the companies. In addition, by separating the shared trend (common market trend) from the company-specific trends, we can better understand how each company performs relative to the overall market dynamics. This helps us distinguish between companies outperforming or underperforming in the market and identify those with unique growth drivers or challenges.

We use Ridge Regression on the sales Data's sum and mean to estimate the shared trend. Ridge Regression is a regularization technique that adds a penalty term to the linear regression model to prevent over-fitting and improve model generalization. The choice between sum and mean for calculating the shared trend depends on the analysis's context and objective. For example, the sum captures the total market trend, influenced by the total sales of all companies combined. In contrast, the mean captures the average market trend, normalizing the effect of different company sizes.

\begin{figure}[t]
\centering
\includegraphics[width=12cm]{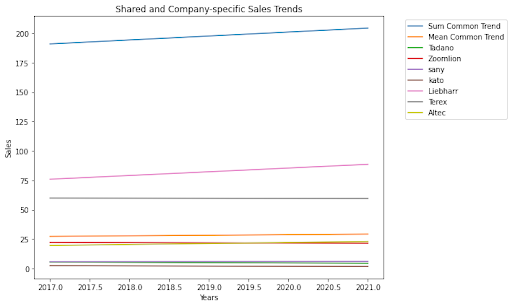}
\caption{\centering Line chart showing the trend analysis for each company
}
\label{fig:4.2}
\end{figure}

Mathematically, the shared trend can be represented as:
\\
For sum:

$\text{shared\_trend\_sum}(t) = a_{\text{sum}} \cdot t + b_{\text{sum}}$
\\
For mean:

$\text{shared\_trend\_avg}(t) = a_{\text{avg}} \cdot t + b_{\text{avg}}$
\\
Where t is time (in years), $a_{\text{sum}}$ and $a_{\text{avg}}$ are the slopes of the shared trend lines for sum and mean, respectively, and $b_{\text{sum}}$ and $b_{\text{avg}}$ are the biases for them. Figure \ref{fig:4.2} is the plot for the trend of the seven companies and the shared direction.

Figure \ref{fig:4.2} is the plot for the trend of the seven companies along with the shared trend
Analyzing these trends allows us to gain insight into each company's performance and how common factors influence it.

The Sum Common Trend and Mean Common Trend represent the overall shared trend in the industry, which reflects the influence of factors such as market demand, economic conditions, and regulatory changes. From 2017 to 2021, both the Sum Common Trend and Mean Common Trend increased steadily, indicating a general growth in the industry during this period.

Focusing on the company-specific trends, we can observe that  Liebherr exhibited a solid upward trend in sales, indicating that the company is performing well and gaining market share. This growth can be attributed to successful product launches, effective marketing strategies, or expansion into new markets.

In conclusion, analyzing the company-specific and shared trends provides valuable insights into each company's performance within the industry. Companies like Sany, Liebherr, and Altec are showing strong growth, while others such as Tadano, Zoomlion, and Kato are experiencing declining trends or facing challenges in maintaining their market share. Terex exhibits a stable trend, indicating that the company is strengthening its position in the market but growing at a different pace than the overall industry. Understanding these trends can help stakeholders make informed investment decisions, business strategies, and potential collaborations.

%% file: Benchmark.tex
\section{Bechmark product simulation and risk factors extraction}
\subsection{Rader chart design benchmark product}

From the above data exploration, it is clear that Liebherr crawler cranes offer several design advantages that make them popular in the construction and heavy-lifting industries, such as:

\textit{High lifting capacity}: Liebherr crawler cranes are designed to handle heavy loads, with lifting capacities ranging from 50 to 3,000 tonnes. 

\textit{Excellent stability}: The crawler tracks on a Liebherr crane provide superior strength and allow the crane to operate on various terrains.

\textit{Versatile boom configurations}: Liebherr crawler cranes can have various boom lengths and attachments, including luffing jibs and fixed or lattice booms. 

\textit{Ease of transportation}: Despite their large size and lifting capacity, Liebherr crawler cranes are designed to transport effortlessly from one job site to another. 

\textit{Advanced technology}: Liebherr is known for its advanced crane technology, including computerized controls, load monitoring systems, and remote monitoring capabilities.
\subsubsection{Simulation benchmark product dimensions}
\textbf{Lifting Capacity}: 

Crane A: "High" lifting capacity compared to other cranes in the benchmark. 

Crane B: "Moderate to High" lifting capacity. 

Crane C: "Moderate" lifting capacity. 

Crane D: "High" lifting capacity. 

Crane E: "Moderate to High" lifting capacity. 

\noindent \textbf{Stability}: 

Crane A: "Excellent" stability during operations. 

Crane B: "Good" stability. 

Crane C: "Very Good" stability. 

Crane D: "Very Good" stability. 

Crane E: "Excellent" stability. 

\noindent \textbf{Boom Configurations}: 

Crane A: "Versatile" boom configurations. 

Crane B: "Highly Versatile" boom configurations. 

Crane C: "Limited" boom configurations. 

Crane D: "Highly Versatile" boom configurations. 

Crane E: "Versatile" boom configurations. 

\noindent \textbf{Transportation Ease}: 

Crane A: "Relatively Easy" to transport compared to other cranes in the benchmark. 

Crane B: "Easy" transportation. 

Crane C: "Very Easy" transportation.

Crane D: "Relatively Easy" transportation. 

Crane E: "Easy" transportation. 

\noindent \textbf{Advanced Technology}: 

Crane A: "Advanced" crane technology, including computerized controls, load monitoring systems, and remote monitoring capabilities. 

Crane B: "Highly Advanced" technology features. 

Crane C: "Moderate" integration of advanced technology. 

Crane D: "Advanced" crane technology. 

Crane E: "Highly Advanced" technology features. 
\\
\\
By using descriptive labels, the simulated data provides a more qualitative representation of the performance or capability of each crane in relation to each dimension. These labels can be further refined or adjusted based on specific industry standards, technical specifications, or expert opinions to accurately reflect the characteristics of each crane in the benchmark. Table1 simulated descriptive labels for the five dimensions mentioned.

\begin{table}[H]
\centering
%\captionsetup{justification=centering}
\caption{\centering Crane Comparison}
\label{tab:crane_comparison}
\begin{tabular}{lc}
\hline
 & \textbf{Values}\\
\hline
\textbf{Lifting Capacity} & \\
- \textbf{Crane A} & High \\
- \textbf{Crane B} & Moderate to High \\
- \textbf{Crane C} & Moderate \\
- \textbf{Crane D} & High \\
- \textbf{Crane E} & Moderate to High \\
\hline
\textbf{Stability} & \\
- \textbf{Crane A} & Excellent \\
- \textbf{Crane B} & Good \\
- \textbf{Crane C} & Very Good \\
- \textbf{Crane D} & Very Good \\
- \textbf{Crane E} & Excellent \\
\hline
\textbf{Boom Configurations} & \\
- \textbf{Crane A} & Versatile \\
- \textbf{Crane B} & Highly Versatile \\
- \textbf{Crane C} & Limited \\
- \textbf{Crane D} & Highly Versatile \\
- \textbf{Crane E} & Versatile \\
\hline
\textbf{Transportation Ease} & \\
- \textbf{Crane A} & Relatively Easy \\
- \textbf{Crane B} & Easy \\
- \textbf{Crane C} & Very Easy \\
- \textbf{Crane D} & Relatively Easy \\
- \textbf{Crane E} & Easy \\
\hline
\textbf{Advanced Technology} & \\
- \textbf{Crane A} & Advanced \\
- \textbf{Crane B} & Highly Advanced \\
- \textbf{Crane C} & Moderate \\
- \textbf{Crane D} & Highly Advanced \\
- \textbf{Crane E} & Highly Advanced \\
\hline
\end{tabular}
\end{table}

\begin{figure}[t]
\centering
\includegraphics[width=12cm]{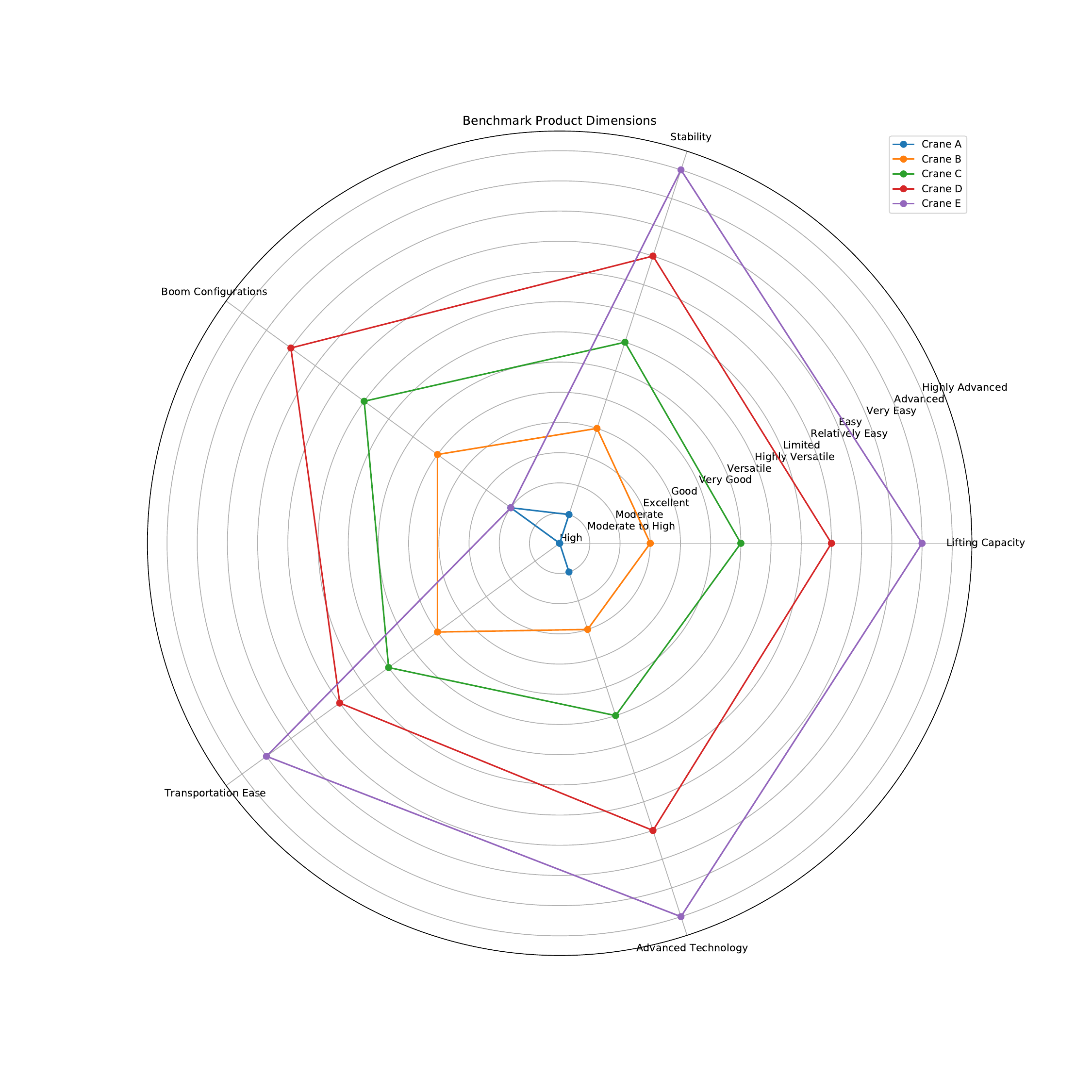}
\caption{\centering A radar chart comparing the different dimensions of the Benchmark product visually
}
\label{fig:radar.png}
\end{figure}

Using this simulated data, Figure \ref{fig:radar.png} shows the radar chart plot to visually compare the Benchmark product's different dimensions. The radar chart will show the performance or capability of each crane about each size, providing a comprehensive view of their strengths and weaknesses in other areas.

\subsection{Risk factors extraction}
According to McGahan (1999) $36\%$ of the variance in profitability could be attributed to the firms' characteristics and actions (\cite{43}). Since all the companies are US local companies with the same political, financial, and marketing environments, the firm its own characteristics and actions are the key factors to distinguish the products sales performances (\cite{44}). 
Besides risk factors of manufacturers' revenue, revenue size, revenue growth rate, and competition relationship analyzed above because of data availability (\cite{45}).  There are some other factors, such as, Product portfolio: The range of products and services offered by manufacturers, including the specifications, features, and performance of heavy machinery equipment (\cite{46}); Market share: The percentage of market share held by each manufacturer in the industry (\cite{47}); Distribution channels: The network of dealers, distributors (\cite{48}); Brand image and reputation: The perception of the brand by customers and stakeholders (\cite{49}); Pricing strategies: The pricing of heavy machinery equipment (\cite{50}); Technological innovation, such as, automation, artificial intelligence, and telematics, to improve the performance and efficiency of heavy machinery equipment. External risk factors, such as, economic downturns, trade policies, and geopolitical tensions (\cite{51,52,53}).
Since too many facotors known or unknown, we devided all risk factors into two categories: \textbf{Internal Growth Potential  and External Competition Dynamics}

%% file: Stochastic.tex
\section{Markov Chain approach for risk factors analysis}
The relationships among these factors can be complex and interdependent. For example, a company's product portfolio can impact its market share, brand image, and pricing strategies. Technological innovation can also impact a company's product portfolio and brand image. External risk factors, such as economic downturns or trade policies, can affect a company's distribution channels, pricing strategies, and overall competitiveness in the market. Therefore, it is important to consider all these factors in a holistic manner when analyzing the competitiveness of the North America Crawler Cranes market. The proposed approach, which utilizes clustering analysis and correlation analysis of revenue data, can provide insights into these complex relationships and help inform marketing decision-making for heavy machinery products.

\subsection{Markov process}
Our investigation of the North American Crawler Cranes market employs a Markov chain model to dissect the dynamics of company performance and market competition. The Markov chain, with its property of memorylessness ($P(X_{n+1}=x|X_1=x_1, X_2=x_2,...,X_n=x_n) = P(X_{n+1}=x|X_n=x_n)$ for all n and $x_1, x_2,...,x_n, x$), characterizes that future state only depends on the current state and not on the sequence of preceding states. This property makes it an apt tool for modeling the rapidly evolving and complex market conditions.
\\
\\
\textbf{Identifying the states}: The first step is to determine appropriate states for the Markov chain. For the company's performance, we define three states, namely 'Declining', 'Stable', and 'Growing'. Similarly, for market competition, we establish three states - 'Declining Competition', 'Stable Competition', and 'Growing Competition'. Hence, we have two sets of states, S = {s1, s2, s3} for the company's performance, and S' = {s1', s2', s3'} for the market competition.
\\
\\
\textbf{Defining transition probabilities and constructing the transition matrices}: For both the company's performance and market competition, we define a transition probability matrix, $P = p_{ij}$ and $P' = p_{ij}'$, where $p_{ij}$ and $p_{ij}'$ represent the probabilities of transitioning from state $s_i$ to state $s_j$ and from state $s_i'$ to state $s_j'$ respectively. These probabilities are computed based on historical sales data. Each row in P and P' is normalized such that $\sum_{j} p_{ij} = 1 \quad \text{and} \quad \sum_{j} p_{ij}' = 1$, satisfying the necessary condition of a stochastic matrix.
\\
\\
\textbf{Calculating the stationary distributions}: The stationary distribution of the Markov chains for the company's performance and market competition provides insights into the long-term behavior of the system. These are probability vectors $\pi= [\pi_1, \pi_2, \pi_3] \quad \text{and} \quad \pi' = [\pi'_1, \pi'_2, \pi'_3]$ that satisfy the conditions $\pi P = \pi \quad \text{and} \quad \pi' P' = \pi'$ respectively. The stationary distributions can be computed using power method or eigenvalue-eigenvector approach. Mathematically, $\pi$ is the left eigenvector of P related to the eigenvalue 1, i.e., $\pi P = \pi$, and similarly for $\pi'$.
\\
\\
\textbf{Analyzing the results}: The stationary distributions offer a measurable estimate of a company's long-term growth potential (internal performance) and the long-term trend in market competition. The stacked bar charts provide a visual comparison of these two aspects for each company, making it a powerful tool for understanding the relative positions of different companies in the market. The comparison of the internal growth potential with the external competition growth potential using the same color scheme in the stacked bar charts offers an intuitive understanding of whether a company is keeping pace with the market competition or not.

This Markov chain-based approach allows us to provide a robust and insightful analysis of the North American Crawler Cranes market, thereby informing strategic decision-making processes.
\begin{figure}[t]
\centering
\includegraphics[width=12cm]{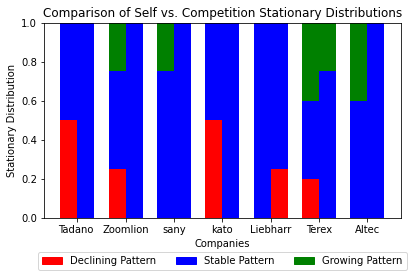}
\caption{\centering Stacked Bar Charts Analyzing Each Company’s Internal Growth and External Competition Environment
}
\label{fig:4.1}
\end{figure}

Figure \ref{fig:4.1} presents the stationary distributions derived from both internal growth models and external competition dynamics. For each company, the first and second bars, moving from left to right, signify the distribution probabilities for internal growth potential and external competition respectively. The color coding used - red, blue, and green - represent declining, stable, and growing states, respectively.

Upon detailed scrutiny of these distributions, we observe:

\textbf{1. Internal Growth Potential}: Tadano and Kato exhibit the highest long-term probability of internal decline. In stark contrast, Sany and Altec display a promising scenario, with substantial probabilities pointing towards internal growth and no indication of potential decline. Terex, while demonstrating a growing trend, does carry some probability of experiencing a decline.

\textbf{2. External Competition Dynamics}: With the exception of Terex, all companies show a tendency towards a stable competition environment. Terex, however, is likely to encounter a growing competitive landscape.

In conclusion, considering both internal growth potential and external competition dynamics, Sany and Altec surface as strong contenders. They exhibit robust internal growth potential, and the likelihood of facing increased competition does not appear to be escalating. This rigorous analysis, based on Markov chain models, serves as a valuable asset for strategic planning and decision-making in the North American Crawler Cranes market.

%% file: Conclusion.tex
\section{Conclusion and future study}

\subsection{Future study on Brownian motion}
Brownian motion, also known as the Wiener process, is a stochastic process commonly used in finance and economics to model random fluctuation over time. It can also be used to model the relationships among the risk factors affecting the crawler crane market.
The Brownian motion model can be expressed mathematically as:
\begin{align}
    dX_t = \mu dt + \sigma dW_t
\end{align}

Where $dX_t$ is the tiny change in the value of factor X at time t, $\mu$ is the drift parameter representing the trend or average rate of change of X over time, $\sigma$ is the volatility parameter representing the randomness or uncertainty of X over time, $dt$ is a tiny time increment, and $dW_t$ is the Wiener process or Brownian motion, which is a stochastic process that represents the random shocks or deviations from the trend.
To model the relationships among the risk factors affecting the crawler crane market using Brownian motion, we can express each element as a separate stochastic process with its own drift and volatility parameters and use correlation coefficients to represent the degree of interdependence. The joint dynamics of the components can then be expressed as a system of stochastic differential equations.
For example, let $X_1, X_2, X_3$ be the stochastic processes representing market share, pricing strategies, and technological innovation, respectively, and let $p_12, p_13, \text{and} p_23$ be the correlation coefficients representing the interdependence among the factors. Then, the joint dynamics of the factors can be expressed as:

\begin{align}
dX_1 &= \mu_1 dt + \sigma_1 dW_1 \\
dX_2 &= \mu_2 dt + \sigma_2 dW_2 \\
dX_3 &= \mu_3 dt + \sigma_3 dW_3
\end{align}

Where $dW_1, dW_2, and dW_3$ are independent Wiener processes, and the drift and volatility parameters $\mu_i$ and $\sigma_i$ are given by:

\begin{align}
\mu_1 &= \alpha_1 + \beta_{12} X_2 + \beta_{13} X_3 \\
\mu_2 &= \alpha_2 + \beta_{21} X_1 + \beta_{23} X_3 \\
\mu_3 &= \alpha_3 + \beta_{31} X_1 + \beta_{32} X_2
\end{align}

$\sigma_1, \sigma_2, \text{and} \sigma_3$ are constants representing the volatility of each factor, and $\alpha_i, \beta_ij$ are parameters that can be estimated from historical data.
This model allows us to simulate possible scenarios and estimate the associated risks by generating random paths for the factors based on the joint distribution of the Wiener processes and the correlation coefficients.

\subsection{Conclusion}
This paper presents a comprehensive approach to competitiveness analysis and risk factor extraction within industrial management and data systems, specifically focusing on the crawler crane market. Competitive analysis is pivotal in industrial management as it involves assessing key market players' strengths and weaknesses, including revenue, revenue size, growth rate, and competitive dynamics.

Industrial companies can gain valuable insights into market trends, customer demands, and competitor strategies by conducting a thorough competitive analysis. These insights inform decision-making, enabling businesses to maintain their competitive edge. They provide actionable data to identify areas where improvements in products, services, or marketing strategies can be made to outperform competitors.

Furthermore, competitive analysis aids in recognizing potential market threats and opportunities. It helps companies develop effective strategies to address these challenges. This may involve identifying unmet market needs, evaluating marketing strategies' effectiveness, and devising innovative products or services to cater to evolving customer requirements.

Moreover, this paper outlines various risk factors relevant to industrial management and data systems competitive analysis. These factors are a foundation for further research and in-depth analysis, facilitating a deeper understanding of the market.

The integration of Markov processes is particularly significant in this field. It enables the capture of the dynamic nature of risk factors and their impact on the market, thereby enhancing informed decision-making. By incorporating Markov processes, industrial companies can gain a more comprehensive understanding of market dynamics, bolster their competitive positioning, and effectively mitigate potential risks.

%% file: main.bbl
\begin{thebibliography}{50}
\newcommand{\enquote}[1]{``#1''}
\providecommand{\natexlab}[1]{#1}
\providecommand{\url}[1]{\normalfont{#1}}
\providecommand{\urlprefix}{}

\bibitem[Adner(2002)]{20}
Adner, Ron. 2002. ``When are technologies disruptive? A demand-based view of
  the emergence of competition.'' \emph{Strategic management journal} 23 (8):
  667--688.

\bibitem[Ahn et~al.(2012)]{61}
Ahn, Jae~Joon, Hyun~Woo Byun, Kyong~Joo Oh, and Tae~Yoon Kim. 2012. ``Using
  ridge regression with genetic algorithm to enhance real estate appraisal
  forecasting.'' \emph{Expert Systems with Applications} 39 (9): 8369--8379.

\bibitem[Asavathiratham(2001)]{6}
Asavathiratham, Chalee. 2001. ``The influence model: A tractable representation
  for the dynamics of networked markov chains.'' PhD diss., Massachusetts
  Institute of Technology.

\bibitem[Ayers and Cordell(2010)]{57}
Ayers, Kristin~L, and Heather~J Cordell. 2010. ``SNP selection in genome-wide
  and candidate gene studies via penalized logistic regression.'' \emph{Genetic
  epidemiology} 34 (8): 879--891.

\bibitem[Barney(1995)]{10}
Barney, Jay~B. 1995. ``Looking inside for competitive advantage.''
  \emph{Academy of Management Perspectives} 9 (4): 49--61.

\bibitem[Bautista-Rodriguez et~al.(2021)]{65}
Bautista-Rodriguez, Carles, Joan Sanchez-de Toledo, Bradley~C Clark, Jethro
  Herberg, Fanny Bajolle, Paula~C Randanne, Diana Salas-Mera, et~al. 2021.
  ``Multisystem inflammatory syndrome in children: an international survey.''
  \emph{Pediatrics} 147 (2).

\bibitem[Bengtsson and Kock(1999)]{22}
Bengtsson, Maria, and S{\"o}ren Kock. 1999. ``Cooperation and competition in
  relationships between competitors in business networks.'' \emph{Journal of
  business \& industrial marketing} 14 (3): 178--194.

\bibitem[Bloodgood and Bauerschmidt(2002)]{9}
Bloodgood, James~M, and Alan Bauerschmidt. 2002. ``Competitive analysis: do
  managers accurately compare their firms to competitors?'' \emph{Journal of
  Managerial Issues} 418--434.

\bibitem[Chan and Ip(2011)]{7}
Chan, SL, and WH~Ip. 2011. ``A dynamic decision support system to predict the
  value of customer for new product development.'' \emph{Decision support
  systems} 52 (1): 178--188.

\bibitem[Chang et~al.(2010)]{62}
Chang, Yu-Chun, Chi-Jui Chang, Kuan-Ta Chen, and Chin-Laung Lei. 2010. ``Radar
  chart: Scanning for high QoE in QoS dimensions.'' In \emph{2010 IEEE
  International Workshop Technical Committee on Communications Quality and
  Reliability (CQR 2010)}, 1--6. IEEE.

\bibitem[Chang et~al.(2012)]{58}
Chang, Yu-Chun, Chi-Jui Chang, Kuan-Ta Chen, and Chin-Laung Lei. 2012. ``Radar
  chart: Scanning for satisfactory QoE in QoS dimensions.'' \emph{IEEE Network}
  26 (4): 25--31.

\bibitem[Chen and Cheng(2007)]{18}
Chen, You-Shyang, and Ching-Hsue Cheng. 2007. ``Forecasting revenue growth rate
  using fundamental analysis: a feature selection based rough sets approach.''
  In \emph{Fourth international conference on fuzzy systems and knowledge
  discovery (FSKD 2007)}, Vol.~3, 151--155. IEEE.

\bibitem[Chinowsky and Meredith(2000)]{11}
Chinowsky, Paul~S, and James~E Meredith. 2000. ``Strategic management in
  construction.'' \emph{Journal of Construction Engineering and Management} 126
  (1): 1--9.

\bibitem[Davies(1973)]{44}
Davies, RL. 1973. ``Evaluation of retail store attributes and sales
  performance.'' \emph{European Journal of Marketing} 7 (2): 89--102.

\bibitem[Deng and Brown(2021)]{52}
Deng, Jiahao, and Eli~T Brown. 2021. ``RISSAD: rule-based interactive
  semi-supervised anomaly detection.'' \emph{Proceedings of EuroVis 2021 short
  papers} .

\bibitem[Dormann et~al.(2013)]{53}
Dormann, Carsten~F, Jane Elith, Sven Bacher, Carsten Buchmann, Gudrun Carl,
  Gabriel Carr{\'e}, Jaime R~Garc{\'\i}a Marqu{\'e}z, et~al. 2013.
  ``Collinearity: a review of methods to deal with it and a simulation study
  evaluating their performance.'' \emph{Ecography} 36 (1): 27--46.

\bibitem[Duan et~al.(2023)]{66}
Duan, Rui, Jiayi Tong, Alex~J Sutton, David~A Asch, Haitao Chu, Christopher~H
  Schmid, and Yong Chen. 2023. ``Origami plot: a novel multivariate data
  visualization tool that improves radar chart.'' \emph{Journal of Clinical
  Epidemiology} 156: 85--94.

\bibitem[Elahi(2013)]{21}
Elahi, Ehsan. 2013. ``Risk management: the next source of competitive
  advantage.'' \emph{Foresight} 15 (2): 117--131.

\bibitem[Harstad(1990)]{15}
Harstad, Ronald~M. 1990. ``Alternative common-value auction procedures: Revenue
  comparisons with free entry.'' \emph{Journal of political economy} 98 (2):
  421--429.

\bibitem[Hartmann(2003)]{17}
Hartmann, George~C. 2003. ``Linking R\&D spending to revenue growth.''
  \emph{Research-Technology Management} 46 (1): 39--46.

\bibitem[Jahan et~al.(2022)]{45}
Jahan, Shah, Khurram Iqbal~Ahmad Khan, Muhammad~Jamaluddin Thaheem, Fahim
  Ullah, Muwaffaq Alqurashi, and Badr~T Alsulami. 2022. ``Modeling
  Profitability-Influencing Risk Factors for Construction Projects: A System
  Dynamics Approach.'' \emph{Buildings} 12 (6): 701.

\bibitem[Jayaprakash et~al.(2016)]{60}
Jayaprakash, Sandeep~M, Eitel~JM Laur{\'\i}a, Pritesh Gandhi, and Dinesh
  Mendhe. 2016. ``Benchmarking student performance and engagement in an early
  alert predictive system using interactive radar charts.'' In
  \emph{Proceedings of the Sixth International Conference on Learning Analytics
  \& Knowledge}, 526--527.

\bibitem[Jugend and Da~Silva(2014)]{46}
Jugend, Daniel, and S{\'e}rgio~Luis Da~Silva. 2014. ``Product-portfolio
  management: A framework based on Methods, Organization, and Strategy.''
  \emph{Concurrent Engineering} 22 (1): 17--28.

\bibitem[Keast(2013)]{50}
Keast, Sarah. 2013. ``Pricing strategy: Setting price levels, managing price
  discounts and establishing price structures.'' .

\bibitem[Koller and Friedman(2009)]{5}
Koller, Daphne, and Nir Friedman. 2009. \emph{Probabilistic graphical models:
  principles and techniques}. MIT press.

\bibitem[Lee et~al.(2016)]{14}
Lee, Kang-Wook, Seung~H Han, Heedae Park, and H~David~Jeong. 2016. ``Empirical
  analysis of host-country effects in the international construction market: An
  industry-level approach.'' \emph{Journal of construction engineering and
  management} 142 (3): 04015092.

\bibitem[Lee et~al.(2011)]{13}
Lee, Sang-Hyo, Rak-Keun Jeon, Ju-Hyung Kim, and Jae-Jun Kim. 2011. ``Strategies
  for developing countries to expand their shares in the global construction
  market: Phase-based SWOT and AAA analyses of Korea.'' \emph{Journal of
  construction engineering and management} 137 (6): 460--470.

\bibitem[L{\'e}vesque, Joglekar, and Davies(2012)]{19}
L{\'e}vesque, Moren, Nitin Joglekar, and Jane Davies. 2012. ``A comparison of
  revenue growth at recent-IPO and established firms: The influence of SG\&A,
  R\&D and COGS.'' \emph{Journal of Business Venturing} 27 (1): 47--61.

\bibitem[Liang et~al.(2023)]{51}
Liang, Yueqing, Canyu Chen, Tian Tian, and Kai Shu. 2023. ``Fair classification
  via domain adaptation: A dual adversarial learning approach.''
  \emph{Frontiers in Big Data} 5: 129.

\bibitem[Lin et~al.(2017)]{59}
Lin, Feng-Han, Sang-Bing Tsai, Yu-Cheng Lee, Cheng-Fu Hsiao, Jie Zhou, Jiangtao
  Wang, and Zhiwen Shang. 2017. ``Empirical research on Kano’s model and
  customer satisfaction.'' \emph{PloS one} 12 (9): e0183888.

\bibitem[Liu et~al.(2008)]{54}
Liu, Wen-Yuan, Bao-Wen Wang, Jia-Xin Yu, Fang Li, Shui-Xing Wang, and Wen-Xue
  Hong. 2008. ``Visualization classification method of multi-dimensional data
  based on radar chart mapping.'' In \emph{2008 International Conference on
  Machine Learning and Cybernetics}, Vol.~2, 857--862. IEEE.

\bibitem[Liu, Crouser, and Ottley(2020)]{64}
Liu, Zhengliang, R~Jordan Crouser, and Alvitta Ottley. 2020. ``Survey on
  individual differences in visualization.'' In \emph{Computer Graphics Forum},
  Vol.~39, 693--712. Wiley Online Library.

\bibitem[McGahan(1999)]{43}
McGahan, Anita~M. 1999. ``The performance of US corporations: 1981--1994.''
  \emph{The Journal of Industrial Economics} 47 (4): 373--398.

\bibitem[Netessine and Shumsky(2005)]{16}
Netessine, Serguei, and Robert~A Shumsky. 2005. ``Revenue management games:
  Horizontal and vertical competition.'' \emph{Management Science} 51 (5):
  813--831.

\bibitem[Nyambane and Bett(2018)]{12}
Nyambane, JM, and S~Bett. 2018. ``Competitive advantage and performance of
  heavy construction equipment suppliers in Kenya: Case of Nairobi County.''
  \emph{International Academic Journal of Human Resource and Business
  Administration} 3 (1): 476--502.

\bibitem[Peng(2022)]{67}
Peng, Weishi. 2022. ``Improved radar chart for lighting system scheme
  selection.'' \emph{Applied Optics} 61 (19): 5619--5625.

\bibitem[Porter, Heppelmann et~al.(2015)]{2}
Porter, Michael~E, James~E Heppelmann, et~al. 2015. ``How smart, connected
  products are transforming companies.'' \emph{Harvard business review} 93
  (10): 96--114.

\bibitem[Prescott and Grant(1988)]{8}
Prescott, John~E, and John~H Grant. 1988. ``A manager's guide for evaluating
  competitive analysis techniques.'' \emph{Interfaces} 18 (3): 10--22.

\bibitem[Putka, Beatty, and Reeder(2018)]{55}
Putka, Dan~J, Adam~S Beatty, and Matthew~C Reeder. 2018. ``Modern prediction
  methods: New perspectives on a common problem.'' \emph{Organizational
  Research Methods} 21 (3): 689--732.

\bibitem[Ranjan(2009)]{1}
Ranjan, Jayanthi. 2009. ``Business intelligence: Concepts, components,
  techniques and benefits.'' \emph{Journal of theoretical and applied
  information technology} 9 (1): 60--70.

\bibitem[Satriadi et~al.(2023)]{69}
Satriadi, Kadek~Ananta, Barrett Ens, Sarah Goodwin, and Tim Dwyer. 2023.
  ``Active Proxy Dashboard: Binding Physical Referents and Abstract Data
  Representations in Situated Visualization through Tangible Interaction.'' In
  \emph{Extended Abstracts of the 2023 CHI Conference on Human Factors in
  Computing Systems}, 1--7.

\bibitem[Smith, Smith, and Wang(2010)]{49}
Smith, Katherine~Taken, Murphy Smith, and Kun Wang. 2010. ``Does brand
  management of corporate reputation translate into higher market value?''
  \emph{Journal of Strategic Marketing} 18 (3): 201--221.

\bibitem[Sornette(2014)]{4}
Sornette, Didier. 2014. ``Physics and financial economics (1776--2014):
  puzzles, Ising and agent-based models.'' \emph{Reports on progress in
  physics} 77 (6): 062001.

\bibitem[Studds(1954)]{63}
Studds, Robert~FA. 1954. ``Radar charting.'' \emph{The International
  Hydrographic Review} .

\bibitem[Trivedi(1998)]{48}
Trivedi, Minakshi. 1998. ``Distribution channels: An extension of exclusive
  retailership.'' \emph{Management science} 44 (7): 896--909.

\bibitem[Tsai et~al.(2015)]{3}
Tsai, Chun-Wei, Chin-Feng Lai, Han-Chieh Chao, and Athanasios~V Vasilakos.
  2015. ``Big data analytics: a survey.'' \emph{Journal of Big data} 2 (1):
  1--32.

\bibitem[Tsao et~al.(2022)]{70}
Tsao, Yen-Po, Wan-Yu Yeh, Teh-Fu Hsu, Lok-Hi Chow, Wei-Chih Chen, Ying-Ying
  Yang, Boaz Shulruf, Chen-Huan Chen, and Hao-Min Cheng. 2022. ``Implementing a
  flipped classroom model in an evidence-based medicine curriculum for
  pre-clinical medical students: evaluating learning effectiveness through
  prospective propensity score-matched cohorts.'' \emph{BMC Medical Education}
  22 (1): 1--11.

\bibitem[Wernerfelt(1986)]{47}
Wernerfelt, Birger. 1986. ``The relation between market share and
  profitability.'' \emph{Journal of Business strategy} 6 (4): 67--74.

\bibitem[Zhang et~al.(2022)]{68}
Zhang, Yongbin, Ronghua Liang, Xiuli Fu, and Yanying Zheng. 2022. ``Visualizing
  Cognitive Learning Outcomes of Undergraduates with Cloud-Based Assessment
  System.'' In \emph{2nd International Conference on Internet, Education and
  Information Technology (IEIT 2022)}, 876--881. Atlantis Press.

\bibitem[Zhang et~al.(2007)]{56}
Zhang, Yu, Zhixiu Hao, Rencheng Wang, and Dewen Jin. 2007. ``A new method for
  the evaluation of gait pathology: A radar chart approach.'' In
  \emph{Proceedings of the 1st international convention on Rehabilitation
  engineering \& assistive technology: in conjunction with 1st Tan Tock Seng
  Hospital Neurorehabilitation Meeting}, 129--132.

\end{thebibliography}
